# Pathway for Multivariate Dependence Modeling in Long-Term Horizon of Electrical Power System

Swasti R. Khuntia, *IEEE Graduate Student Member*
José L. Rueda, *IEEE Senior Member*
Dept. of Electrical Sustainable Energy
Delft University of Technology
Delft, The Netherlands
E-mail: s.r.khuntia@tudelft.nl

Mart A.M.M. van der Meijden[1,2], *IEEE Member*
[1]Dept. of Electrical Sustainable Energy
Delft University of Technology
Delft, The Netherlands
[2]Tennet TSO B.V.
Arnhem, The Netherlands

*Abstract*—Future electricity consumption is fundamentally uncertain and dependent on many variables such as economic activity, weather, electricity rates and demand side management. The stochasticity of system load as well as power generation from renewable energy sources (RES) (i.e., wind and solar) poses special challenges to power system planners. Increasing penetration levels of wind and solar exacerbate the uncertainty and variability that must be addressed in coming years, and can be extremely relevant to power system planners. With this paper, pathways for multivariate dependence modeling using vine copula is proposed which includes both electrical load and power generation from RES.

## I. INTRODUCTION

With the large number of factors affecting long-term decision making process, accuracy of the forecasting method and uncertainty modeling is vital. For example, the lack of accuracy of load forecast modeling can impact big time investment costs and time. As reported in [1], the volatility plays a crucial role in designing an accurate model for long-term load forecast. Long-term horizon spans from a few years (> 1 year) to 10-20 years [2]. In such a long time-frame, modeling of electrical load and power generation from renewable energy sources (RES) is a herculean task for planners because it plays an important role for system planning, scheduling expansion of generation units by construction and procurement of generation units. And, because it takes several years and requires a huge investment for construction of power generation and transmission facilities, accurate and error-free forecasting is necessary for an electric utility. For instance, European project GARPUR (www.garpur-project.eu) is working on this aspect. It is important to introduce the notion of exogenous variables that are modelled explicitly in a reliability management task. Exogenous variables are determined external to the system operators, and the system operators will have to adapt its behavior accordingly [3].

Important question is how it is to be done? *Starting off with electrical load, it is intended to analyze load forecast error time series, and understand it as a stochastic value. The same procedure is followed for modeling wind and solar power as stochastic variables, which may have several possible future realizations.*

In general, approaches for uncertainty modeling include non-probabilistic approaches and probabilistic approaches. For non-probabilistic approaches, interval analysis, fuzzy theory, possibility theory and evidence theory are most widely used. Probabilistic approach is being considered as the most rigorous approach to uncertainty analysis in engineering design due to its consistency with the theory of decision analysis. This is only true if we have a way of quantifying the probability measure implied by the uncertain parameters. The fundamental characterization of probability is the probability density function (PDF). In probability theory, a PDF, or density of a continuous random variable, is a function that describes the relative likelihood for this random variable to take on a given value. The probability for the random variable to fall within a particular region is given by the integral of this variables density over the region. The two main probabilistic approaches that are widely used are analytical methods and the stochastic methods (or the Monte-Carlo (MC) simulation methods). The analytical method requires a number of simplification methods in order to reach an analytical formulation of the problem, namely, linearization of the model, assuming inputs as independent, and normally distributed. Analytical methods are preferred in load flow studies [4] but it has some drawbacks too. For example, assumption of independence of the nodal loads is quite unrealistic. In stochastic methods, the Monte-Carlo (MC) simulation method is the general designation for stochastic simulations using random numbers [5]. Compared to the analytical methods, stochastic methods offer significant advantages, since the basic computational part is deterministic and there is no need to simplify the mathematical models to ensure applicability of the method.

The rest of the paper is organized as follows: Section II discusses load forecast in long-term horizon. Section III presents an understanding of power generation from RES in long-term horizon. Section IV discusses the pathway for multivariate dependency modeling. Finally, section V concludes the work.

## II. Load Forecast in Long-term Horizon

Long-term load forecasting plays an important role in power systems for system planning, scheduling expansion of generation units by construction and procurement of generation units. Because it takes several years and requires a huge investment for construction of power generation and transmission facilities, accurate and error-free forecasting is necessary for an electric utility. Accuracy of load forecast has a direct impact on development of future generation and transmission planning, and hence it is a crucial instrument for planning and forecasting future conditions of the electricity network. Based on the forecast, electric utilities coordinate their resources to meet the forecasted demand using a least-cost plan. In general, forecasting is subjected to a large number of uncertainties and ample amount of research indicates that load predication in presence of uncertainties is required for future capacity resource needs and operation of existing generation resources.

Long-term load forecasting is much more complex than simply fitting a mathematical model to some data, and it requires a lot more knowledge about the "substantive" problem. Compared to short term load forecast that uses a sort of exercise on data modeling (for e.g., fitting models to datasets and extrapolating from them, without really understanding much about the way an electrical system works), long-term load forecast, on the other hand, depends less on the analyst's expertise on modeling, and more on experience with power systems, and a thorough understanding of the way the system works, and how the electricity market may be affected by the changes in a country's economy throughout the years, or by changes in technology, etc.

The load forecast methodology takes into account some explicit factors like historical load and weather data, economic indicators like gross domestic product and their forecasts, and demographic data which includes consumer data like population, appliances in use, etc. In the long run, it has become more important for planners and forecasters to study movements of load time-series and its fluctuations. The movements are usually measured by the volatility (or conditional standard deviation) of the load time-series. One of the biggest problem with modeling the volatility is one of its features, it has periods with low movements and then suddenly periods with high movements. Ref. [1] can be referred to for more discussion on volatility issue in long-term load forecast.

## III. Renewable Energy Sources (RES) Power Forecast in Long-term Horizon

Uncertainty and variability can be seen as the two fundamental characteristics associated to renewable energy sources (RES), specially wind and solar/photo-voltaic (PV) generation, when compared to conventional power generation. The output power of both wind farm and PV plants are vulnerable to sudden weather changes, and it can vary from decent value to even zero. This can significantly impact the normal operation; hence failing to consider RES forecast and its associated uncertainty can translate in a poor decision making, negatively affecting the cost and reliability of the system. There are different timescales involved in the forecasting process: from the evaluation of few seconds-ahead wind variations in order to enhance the efficiency of a wind turbine pitch control strategy, to 20-30 year-ahead resource estimation to be used in planning studies. In such cases, high wind power penetration will have significant impacts on power system operation economics, stability, security, and reliability due to fast fluctuation and unpredictable characteristics of wind speed.

Uncertainty in wind generating sources can be divided as:
- Wind speed distribution
- Uncertain power curve

The uncertainties of wind generating sources are mainly from two aspects: the intermittent and volatile wind speed and the uncertain power curve. Wind speed, as an essential measurement for wind power generation, is influenced by many factors such as the weather conditions, the land terrain, and the height above the ground surface. The power curve of a wind turbine is a graph that indicates the mathematical mapping from different wind speeds to the electrical power output of wind turbines.

PV generation forecasting accuracy is mainly influenced by the variability in meteorological conditions and to a minor extent to the uncertainties related to the different modeling steps needed to predict power generation from meteorological forecasts [6]. Solar generation presents a high sensitivity to changes in solar irradiation, making the variability (probability of ramp events or sudden changes in production) an additional component of a comprehensive solar forecast.

## IV. Pathway for Multivariate Dependency Modeling: Univariate vs. Multivariate

It is important to capture the inherent dependence between load and power generated from RES (i.e., wind and solar generation). In order to explore and exploit the impacts of wind power uncertainty on the power grid whilst considering load uncertainties, a numerical simulation technique like MC-approach is generally applied. However, this creates a heavy computational burden due to the necessarily large sample size. Scenario generation is an important part of uncertainty modeling. It helps in approximating distributions when the number of outcomes are infinite but can perform poorly when using only a few outcomes. In order to simulate scenarios for two or more variables jointly, their correlation structure must be taken into account. For instance, if the studied time-series is Gaussian, then Cholesky decomposition is enough to generate a correlated simulation, otherwise a copula approach is necessary.

It can be said that univariate methods are easier to ensure robustness and considered to be sufficient for short lead times because weather variables like wind and irradiation that affect wind and solar forecast tend to change in a smooth fashion over short time frames, what will be captured in the demand series itself. On the other hand, multivariate time-series use multiple predictors to predict future behavior of one or more response variables [7]. In the long-term, load forecasting is usually approached as a multivariate problem, having multiple predictors and one (in the case of peak load) or more (in the case of load profile) outputs. Ref. [8] can be referred for long-term load forecast. Another well-known theory regarding univariate and multivariate modeling states that multivariate model series can be regarded as a set of univariate models for each of the series. In such case, univariate results can provide a benchmark for multivariate models, and should be explained by them. More generally, however, since the multivariate model implies a set of univariate models, these should be derived from the fitted multivariate one, and then compared to the models obtained through univariate analysis. If the two sets of univariate models are clearly different, then there is reason to suspect specification error in some of the models. For our study, rather than integrating different univariate model series into multivariate model series, we propose a multivariate modeling approach.

The integration of RES in a bulk power system is increasing and will increase more in coming years, and call for the importance of modeling the joint probability distribution. The joint distribution incorporates the one-dimensional marginal distributions of load and wind and these latter are the most easy to assess through data analysis or expert judgement. But, in reality, the one-dimensional marginal distributions are not sufficient to calculate the joint distribution. They are sufficient if load and power generated from RES are independent random variables, but in reality they are not independent. Obtaining the joint distribution given the marginals is however a non-trivial problem, since there exist an infinite number of joint distributions with the same marginals, corresponding to an infinite number of stochastic dependence structures between the random variables load and wind.

The modeling framework can be divided into two major tasks:
- Modeling the one-dimensional marginal distributions called marginals.
- Modeling the stochastic dependence.

It is important to capture these two aspects. For example, if the marginal distributions remain the same, the joint probability distribution can change due to changes in the dependence.

## A. Modeling the one-dimensional marginal distributions

The one-dimension marginal distributions capture the stochastic behavior of the individual exogenous parameters. The marginal cumulative density function (CDF) of a random variable $X$ is defined as

$$CDF_X(x) = P(X \leq x)$$

An important property of CDFs is that the cumulative distribution function of $X$ applied to $X$ itself yields a uniformly distributed random variable. Mathematically, it can be written as [9]:

$$For\ x\epsilon[0,1]: P(CDF_X(X) \leq x) = P(X \leq CDF_X^{-1}(x))$$
$$= CDF_X[CDF_X^{-1}(x)] = x$$

This forms the base of MC-sampling method. Using the equation above, sampling a random variable $X$ with CDF $CDF_X$ can be done by first sampling a random realization $y$ from a uniform random variable $Y$ in [0,1], and then applying the transformation $x = CDF_X^{-1}(y)$. In these cases, the samples $x$ follow the distribution $CDF_X$ [10]. It can be extended for sampling from real measured data by using the empirical CDF. Fig. 1 shows an example of an empirical marginal CDF for the forecast error of one single load.

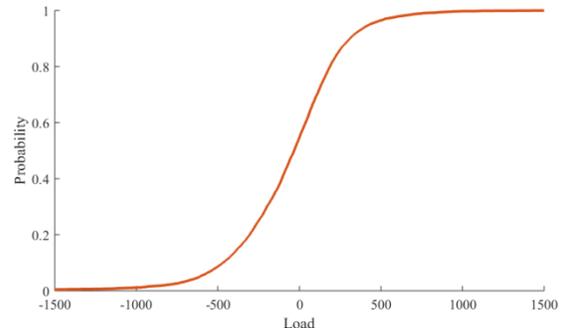

Fig. 1: Empirical marginal CDF of forecast error of a single load.

The sampling method above can be applied for sampling any single random variable whose CDF is known. When several random variables that are correlated are to be sampled, the above is not enough since it does not capture any measure of the dependence between the random variables. However, the marginal CDFs are still important as will be seen in the next section.

## B. Modeling the stochastic dependence

For modeling RES, the dependence between the prime movers (wind speed and solar irradiation) and output power (wind and solar power) is important to study. In case of wind power, the wind turbine generator behaves differently in different geographic locations. In general, the wind resources present significant correlation. The same also applies for solar power (i.e., solar irradiation varies with location change). In particular, the power output of stochastic generators situated in a small geographic area show similar fluctuations due to their mutual dependence on the same prime mover, which is not the case for stochastic generators situated in remote areas. With increased penetration of RES into primary grid, it is vital to model the complex interdependencies introduced by the RES in the power system along with system load.

The dependence between exogenous parameters can be captured by different measures of dependence. For general multivariate random variables, Spearman's rank coefficient [11] can be used to study the non-linear, monotonic relationship between two random variables. Spearman's rank coefficient helps in defining the dependence structure based on rank with specific functions, called copula functions. Using copula functions, it is possible to simulate two random variables that are correlated according to rank correlation by first simulating a copula and later transforming the obtained ranks into respective marginals. Copula is a multivariate probability distribution for which the marginal probability distribution of each variable is uniform [12]. Copulas are used to describe the dependence between random variables. By definition, for two random variables $X$ and $Y$ with CDF $CDF_X, CDF_Y$ are joint by copula $C$ if their joint distribution can be written as:

$$CDF_{XY}(x,y) = C(CDF_X(x), CDF_Y(y))$$

The function $C$ is therefore defined on uniformly random variables, and the cumulative distribution functions can be used to map the uniformly random variables to $X$ and $Y$. Algorithm 1 describes the steps to sample two random variables using copula.

| **Algorithm 1:** Sampling of two correlated random variables using copula |
|---|
| **Inputs:** Two uniform independent random variables $U_{r1}, U_{r2}$ (say load and wind power), correlation coefficient, copula function. |
| **Outputs:** Sampled distributions |
| 1 Sample two uniform random variables $U_{r1}, U_{r2}$, and obtain the realizations $u_{r1}, u_{r2}$. |
| 2 $u_1 = u_{r1}$: presents sampling of the rank distribution $U_1$ |
| 3 Calculate copula function $C_{12|u_1, \rho_{r12}}$, i.e., the conditional distribution of $U_2$ for rank correlation $\rho_{r12}$ and given $u_1$. |
| 4 Sample the rank distribution $U_2$: $u_2 = C^{-1}_{12|u_1, \rho_{r12}}(u_{r2})$, using the inverse copula function. Outcome can be either 0 or constant, depending on the value of independent sample $u_{r2}$ |
| 5 Transform the rank distributions according to the marginals: $x_1 = CDF_1^{-1}(u_1)$ and $x_2 = CDF_2^{-1}(u_2)$ |
| 6 End |

Note that for a given rank coefficient, different copula functions can be used as described in [13]. A particular instance of the above algorithm is the use of a Gaussian copula. When using Gaussian copula, the overall method in Algorithm 1 is called the joint normal transform [14]. The mean of the Gaussian copula is zero and the covariance matrix is

$$R = \begin{pmatrix} 1 & \sigma \\ \sigma & 1 \end{pmatrix}.$$

A property of the Gaussian copula is that the covariance $\sigma$ can be computed from the rank correlation of $X$ and $Y$ as follows:

$$\sigma = 2\sin\left(\frac{\pi}{6}\rho_r(X,Y)\right).$$

Note that $\sigma$ is not the covariance between $X$ and $Y$ but the covariance used in the Gaussian copula. The equation above links this covariance with the rank coefficient of $X$ and $Y$.

The above procedure describes the joint normal transform for two random variables. The procedure can be generalized to $n$ random variables by applying the above to all pairs of random variables. The resulting Gaussian copula will be $n$-dimensional. The application of joint normal transform can be used for any system with following data availability:

- Marginal distributions: The system load distribution, wind speed distributions and solar power generation at each generation node of the system.
- Dependence structure: The product moment correlation matrix is calculated from rank correlation matrix, between all pairs of exogenous parameters.

The above algorithm works well when all required data is available because it is needed to calculate the correlation and then later use copula to model the stochastic dependency. In reality, not all required data is available. The reason can be anything from bad measurement devices to confidential information or just 'stochasticity'. In case of data unavailability, mutual correlations between the stochastic inputs is not possible and hence the joint normal transform is not recommended. To tackle this issue, a vine copula model offers suitable solution [10]. In addition, some shortcomings of multidimensional copula functions and the Gaussian copula favor the use of vine copula models, which is preferred as discussed previously. The shortcomings are studied by [15] and are practical infeasibility and dimensionality issues.

Vine copula models decompose a multivariate copula into a set of bivariate copulas and every bivariate copula function can be imagined as a branch of a graph connecting two consecutive marginal distributions or their conditional bivariate distributions. By definition [10], a vine copula on $n$ variables is a nested set of trees $T_j$ where the edges of the $j^{th}$ tree become the nodes of the $(j+1)^{st}$ tree for $j = 1, \dots, n$. A regular vine on $n$ variables is defined as a vine in which two edges in tree $j$ are joined by an edge in tree $j+1$ only if these edges share a common node. Each edge in the regular vine may be associated with a conditional rank correlation and a copula, and each node with a marginal distribution. All assignments of rank correlations to edges of a vine are consistent and each one of these correlations may be realized by a copula. Based on the bivariate and conditional bivariate distributions, the joint distribution can be constructed. Recent studies have attracted the use of vine copulas and can be found in refs. [16-17].

A regular vine can be either d – vine where each node in $T_j$ has a degree of at most 2 or c (canonical) c – vine in which each tree $T_j$ has a unique node of degree $n - i$. Fig. 2 shows the d-vine on four uniform variables labelled $X_1, X_2, X_3, X_4$.

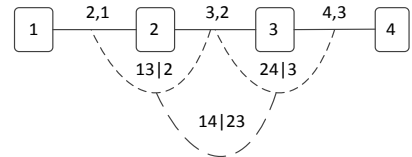

Fig. 2: d-vine on four variables

Distributions specified by conditional rank correlation on a d-vine can be sampled, and an algorithm to do it is presented in Algorithm 2, which can be expanded from 4 to $n$ variables. The algorithm involves sampling four independent uniform (0,1) variables $U_1, U_2, U_3, U_4$. The conditional correlation between variables $(i, j)$ given $k$ is given as $\rho_{r_{ij|k}}$. The $CDF$ for $X_j$ given $U_i$ under the conditional copula with correlation $r_{ij|k}$ is given as $CDF_{\rho_{r_{ij|k};U_i}}(X_j)$.

| | **Algorithm 2:** Sampling of 4 uniform random variables using d-vine |
|---|---|
| | **Inputs:** Four uniform random variables $\{X_1, X_2, X_3, X_4\}$, conditional rank correlations and corresponding CDFs |
| | **Outputs:** Sampled distributions |
| 1 | $x_1 = u_1$. |
| 2 | $x_2 = CDF^{-1}_{\rho_{r12;x_1}}(u_2)$. |
| 3 | $x_3 = CDF^{-1}_{\rho_{r23;x_2}}(CDF^{-1}_{\rho_{r13|2;CDF_{\rho_{r12;x_2}}(X_1)}}(u_3))$. |
| 4 | $x_4 = CDF^{-1}_{\rho_{r34;x_3}}(CDF^{-1}_{\rho_{r24|3;CDF_{\rho_{r23;x_3}}(X_2)}}(CDF^{-1}_{\rho_{r14|23;CDF_{\rho_{r13|2;CDF_{23;x_2}(x_3)}}(CDF_{\rho_{r12;x_2}}(x_1))}}(u_4)))$ |
| 5 | **End** |

## V. Conclusion and Discussion

The pathway described in this paper can be well adapted to sample *n*-dimensional distribution. It can be concluded that for uncertainty modeling, it is recommended to devise a multivariate problem taking load, wind speed and solar irradiation together. The solution follows MC-sampling approach with two basic components: modeling of one-dimensional marginal distributions and modeling of the multi-dimensional stochastic dependence structure. The multivariate normal distribution is measured by product moment correlation, followed by creation of correlation rank matrix. Finally, scenario generation is performed by the use of copula functions. In short, simulation of bivariate relationships between dependent random variables can be performed using the marginal distributions, the rank correlation and the respective copula functions.

Along with the specifications for uncertainty modeling, data requirements for realization of above discussed methodology is important. It can be recalled that data availability is a deciding factor, whether to go for dependency modeling using the joint normal transform or select available features from inadequate data and use dependency modeling using vine copula models.

The selection of an adequate method is restricted by the data availability and the daily grid operation processes. A number of factors influence the uncertainty in load and RES forecasting. First, weather data at the nodal scale is limited, which decreases the accuracy of forecasts for both nodal load and RES production. Second, switching operations in underlying grids could lead to unexpected deviations from the point forecast at certain grid nodes. Third, maintenance activities in wind and solar farms may not be reported, which would lead to an overestimation of the forecast. The latter two factors can only be observed and, thus, incorporated in the uncertainty estimation when real data becomes available.

## References

[1]. S.R. Khuntia, J.L. Rueda, and M.A.M.M. van der Meijden, " Volatility in electrical load forecasting for long-term horizon – An ARIMA-GARCH approach," *Proc. PMAPS 2016*.
[2]. S.R. Khuntia, B.W. Tuinema, J.L. Rueda, and M.A.M.M. van der Meijden, "Time-horizons in the planning and operation of transmission networks: an overview," *IET Gen. Trans. Dist.*, vol. 10, no. 4, pp. 841-848, 2016.
[3]. Guidelines for implementing the new reliability assessment and optimization methodology, GARPUR consortium, http://www.garpur-project.eu/deliverables.
[4]. M. Schilling, A.M. Leite da Silva, R. Billington, and M.A. El-Kady. "Bibliography on power system probabilistic analysis (1962-1988)." *IEEE Trans. Power Syst.* Vol. 5, no. 1, pp. 1-11, 1990.
[5]. R. Rubinstein. *Simulation and the Monte Carlo method*. John Wiley & Sons, New York, 1981.
[6]. S. Pelland, G. Galanis, and G. Kallos, "Solar and photovoltaic forecasting through post-processing of the Global Environmental Multiscale numerical weather prediction model," *Progress Photovoltaics: Res. Appl.*, vol. 21, no. 3, pp. 284-296, 2013.
[7]. G.C. Reinsel. Elements of Multivariate Time Series Analysis. Springer Science & Business Media, 2003.
[8]. S.R. Khuntia, J.L. Rueda, and M.A.M.M. van der Meijden, "Forecasting the load of electrical power systems in mid- and long-term horizons - A review," *IET Gen. Trans Dist.*, In press.
[9]. D. Kurowicka and R.M. Cooke. *Uncertainty analysis with high dimensional dependence modelling*. Wiley series in Probability and Statistics, 2006.
[10]. D. Kurowicka. *Dependence modeling: vine copula handbook*. World Scientific, 2011.
[11]. P. Embrechts, F. Lindskog, and A. McNeil, Modelling dependence with copulas. *Rapport technique, Département de mathématiques, Institut Fédéral de Technologie de Zurich, Zurich*, 2001.
[12]. R.B. Nelsen, An Introduction to Copulas. New York, NY, USA: Springer, 2006.
[13]. H. Louie, "Evaluation of bivariate Archimedean and elliptical copulas to model wind power dependency structures," *Wind Energ.*, vol. 17, no. 2, pp. 225–240, 2014.
[14]. G. Papaefthymiou, and D. Kurowicka, "Using copulas for modeling stochastic dependence in power system uncertainty analysis," *IEEE Trans. Power Syst.*, vol. 24, no.1, pp. 40-49, 2009.
[15]. T. Bedford and R.M. Cooke, "Vines—A new graphical model for dependent random variables," *Ann. Statist.*, vol. 30, no. 4, pp. 1031–1068, 2002.
[16]. R.J. Bessa, "On the quality of the Gaussian copula for multi-temporal decision-making problems," *Proc. PSCC 2016*.
[17]. M. Sun, et al., "Evaluating composite approaches to modelling high-dimensional stochastic variables in power systems," *Proc. PSCC 2016*.